\documentclass[sigconf,nonacm,prologue,table]{acmart}

\AtBeginDocument{%
  \providecommand\BibTeX{{%
    \normalfont B\kern-0.5em{\scshape i\kern-0.25em b}\kern-0.8em\TeX}}}
\usepackage[T1]{fontenc}

\usepackage{adjustbox}

\usepackage[justification=centering]{caption}

\usepackage{graphicx}
\usepackage{subfig}
\graphicspath{ {./images/} }

\usepackage{algorithm}
\usepackage{algpseudocode}

\usepackage{amsmath}

\usepackage{listings}

\definecolor{lightgray}{rgb}{.9,.9,.9}
\definecolor{darkgray}{rgb}{.4,.4,.4}
\definecolor{purple}{rgb}{0.65, 0.12, 0.82}

\lstdefinelanguage{JavaScript}{
  keywords={typeof, new, true, false, catch, function, return, null, catch, switch, var, if, in, while, do, else, case, break},
  keywordstyle=\color{blue}\bfseries,
  ndkeywords={class, export, boolean, throw, implements, import, this},
  ndkeywordstyle=\color{darkgray}\bfseries,
  identifierstyle=\color{black},
  sensitive=false,
  comment=[l]{//},
  morecomment=[s]{/*}{*/},
  commentstyle=\color{purple}\ttfamily,
  stringstyle=\color{red}\ttfamily,
  morestring=[b]',
  morestring=[b]"
}

\usepackage{xspace}
\newcommand{\radar}[0]{\textsc{UA-Radar}\xspace}

\begin{document}

\author{Jean Luc Intumwayase}
\email{intumwayase@gmail.com}
\orcid{0009-0002-0525-1565}
\affiliation{%
  \institution{Univ. Lille, Inria, CNRS, UMR CRIStAL 9189}
   \city{Lille}
   \country{France}
}

\author{Imane Fouad}
\email{imane.fouad@inria.fr}
\orcid{0009-0006-7613-0428}
\affiliation{%
  \institution{Univ. Lille, Inria, CNRS, UMR CRIStAL 9189}
   \city{Lille}
   \country{France}
}

\author{Pierre Laperdrix}
\email{pierre.laperdrix@inria.fr}
\orcid{0000-0001-6901-3596}
\affiliation{%
  \institution{Univ. Lille, Inria, CNRS, UMR CRIStAL 9189}
   \city{Lille}
   \country{France}
}

\author{Romain Rouvoy}
\email{romain.rouvoy@univ-lille.fr}
\orcid{0000-0003-1771-8791}
\affiliation{%
  \institution{Univ. Lille, Inria, CNRS, UMR CRIStAL 9189}
   \city{Lille}
   \country{France}
}


\title{\radar: Exploring the Impact of User Agents on the Web}

\begin{abstract}
  In the early days of the web, giving the same web page to different browsers could provide very different results. 
  As the rendering engine behind each browser would differ, some elements of a page could break or be positioned in the wrong location. 
  At that time, the \emph{User Agent} (UA) string was introduced for content negotiation.
  By knowing the browser used to connect to the server, a developer could provide a web page that was tailored for that specific browser to remove any usability problems.
  Over the past three decades, the UA string remained exposed by browsers, but its current usefulness is being debated. 
  Browsers now adopt the exact same standards and use the same languages to display the same content to users, bringing the question if the content of the UA string is still relevant today, or if it is a relic of the past.
  Moreover, the diversity of means to browse the web has become so large that the UA string is one of the top contributors to tracking users in the field of browser fingerprinting, bringing a sense of urgency to deprecate it.
  
  In this paper, our goal is to understand the impact of the UA on the web and if this legacy string is still actively used to adapt the content served to users.
  We introduce \radar, a web page similarity measurement tool that compares in-depth two web pages from the code to their actual rendering, and highlights the similarities it finds.
  We crawled $270,048$ web pages from $11,252$ domains using $3$ different browsers and $2$ different UA strings to observe that 100\% of the web pages were similar before any JavaScript was executed, demonstrating the absence of differential serving.
  Our experiments also show that only a very small number of websites are affected by the lack of UA information, which can be fixed in most cases by updating code to become browser-agnostic. 
  Our study brings some proof that it may be time to turn the page on the UA string and retire it from current web browsers.
\end{abstract}

\maketitle

\section{Introduction}\label{introduction}
In the early days of the web, web browsers had different technological stacks and would not interpret HTML tags the exact same way~\cite{grosskurth_architecture_nodate}.
This created usability problems as the exact same version of a web page would render differently on different browsers.
To remedy this problem, each browser started to include a \emph{User Agent} (UA) header that would expose the browser and its version to the server.
Web developers could then provide a version that was tailored to the user's browser so that the website would appear as intended with all the elements in the right place.

In 2023, more than 30 years after it was first officially introduced~\cite{uaStandard}, the UA string is still being used and its history is long, granular, and complex~\cite{kline_structure_2017}. 
What was first introduced as a tool to help servers to deliver the most optimized content to users became a source of competition and now tracking~\cite{uaHistory}.
In particular, UA exposed by browsers can be leveraged by browser fingerprinting, which has seen a steady rise in the past decade~\cite{laperdrix_browser_2020}. 
By running a little script on a web page, a server can collect a wide range of information on the device being used by the user from the browser and its version to the size of the screen or the GPU.
The diversity of today's devices and configurations is so large that it is possible to identify users based only on this information. 
No other identifiers like cookies are needed to track users on the Internet if a fingerprint is precise enough. 
Because of the danger posed by fingerprinting, some browser vendors started to make modifications to limit the information revealed by the browser. 
One such initiative is the UA Client Hints by Google~\cite{ua_reduction}, whose goal is to freeze the UA string as it is one of the most revealing information in fingerprints~\cite{eckersley_how_2010,laperdrix16Beauty}.

In this paper, we investigate the impact of the UA string on the web and whether servers still leverage it to adapt the content that is served to users.
We introduce \radar, a web page similarity measurement tool to assess the impact of restricting the User-Agent request-header field, the \texttt{navigator.userAgent}, and other identifying information in the Navigator object on the web.
We crawl $270,048$ web pages from $11,252$ domains using standard and so-called \emph{none} browsers, and we observe 100\% similarity of the web pages before the execution of JavaScript, demonstrating the absence of differential serving.
However, 8.4\% of the web pages change after the execution of JavaScript, hence highlighting dependency on UA for content adaptation.
We conduct a change impact analysis on UA-dependent web pages and find third-party scripts from ads, bot detection, and content delivery network services behind the changes in the web pages.

This paper addresses the following research questions:
\begin{itemize}
\item {\bf RQ1}: Do modern websites adapt to the UA?
\item {\bf RQ2}: What are the changes created by different UA? What are their causes?
\item {\bf RQ3}: What is the impact of removing identifying information from the UA?
\end{itemize}

The remainder of this paper is organized as follows.
Section~\ref{related_work} describes the related work.
Section~\ref{methodology} details our methodology and how we measure web similarity in the wild.
Section~\ref{dataset} goes over details of our crawl and analysis of our dataset.
Section~\ref{discussion} discusses the impact of UA on the web, based on our findings.
Section~\ref{threats_to_validity} discusses the threats to the validity of our work and Section~\ref{conclusion} concludes the paper.

\section{Related Work}\label{related_work}
\subsection{User Agent on the Web}\label{subsec:uaweb}
In the literature, UA has been studied for security improvements, network monitoring, Internet traffic analysis, and user behavior analysis~\cite{choudhary_detecting_2011, kheir_behavioral_2013, kline_structure_2017, nejati_2016, la_network_2016, pham_understanding_2016, islam_user_agent}.
We want to highlight here 2 specific areas that are relevant to this study.

\paragraph{\textbf{The threat of the UA to privacy.}}
In the past decade, a tracking technique known as browser fingerprinting~\cite{laperdrix_browser_2020} has grown a lot. 
By collecting attributes in a web browser from HTTP headers and JavaScript, a script can build the \textit{fingerprint} of a device that can be used later on to identify a user~\cite{eckersley_how_2010}. 
The diversity of hardware and software configuration is so broad that the combination of all collected attributes can be unique.
According to three large-scale fingerprinting studies~\cite{eckersley_how_2010,gomez18Hiding,laperdrix16Beauty}, the HTTP user agent is one of the most revealing attributes in a fingerprint, as it always ended up in the top 3 of collected attributes with the most entropy. 
In particular, this header on mobile can even reveal the exact model of the user's smartphone along with the phone carrier which is very worrisome for privacy~\cite{laperdrix16Beauty}. 
For this reason, we want to investigate in this study how useful the UA is on today's web, more than 25 years after it was introduced, because if it is possible to remove it, it would seriously affect the capacity of browser fingerprinting to track users online.

\paragraph{\textbf{Restricting the User Agent.}}
In the past, developers of the Safari browser froze the UA to reduce web compatibility risks and to prevent the use of UA for browser fingerprinting~\cite{safari_release_2017}.
The community quickly reported page breakages, due to lack of UA information.
Building on past experience, the W3C Community Group introduced UA client hints to reduce UA granularity~\cite{ua_reduction}.
Developers of Chrome browser built on that work to implement UA reduction and has since been progressively released in new Chrome versions~\cite{ua_reduction}.
Developers at Mozilla/Tor have been working on browser fingerprint resistance and have collected several reports of web page breakages~\cite{fingerprint_bug}.
However, no study has been conducted to study the impact of frozen or reduced UA on the web.

\subsection{Measuring Web Page Similarities}
Web similarity was previously studied for content categorization, anti-phishing, browser performance optimization, user experience, web archiving, and crawling strategy~\cite{strehl_impact_nodate, wenyin_detection_2005, wang_similarity-based_2014, chen_detecting_2003, law_structural_2012, ali_whats_2003, eravuchira_measuring_2016}.
Those studies were limited to individual comparisons of text, visual rendering, or HTML structure.
Tombros~\emph{et al.} studied factors that determine web page similarity by evaluating the effectiveness of HTML structure and content~\cite{tombros_factors_2005}. Hashmi et al. worked on QLUE, a visual comparison tool that evaluates the content and functionality of web pages using structural similarity~\cite{hashmi_qlue_2022}. QLUE takes an excessive amount of time to produce results, hindering its scalability, so a novel approach was required to compare visual rendering in the wild. No work has been done before to develop compound metrics to fully understand the similarity of web pages. Our paper is the first that provides similarity metrics to explore the impact of UA on the web.

\section{\radar: Measuring Web Similarity in the Wild}\label{methodology}

\subsection{Overview}
Given the evolution of web technologies, measuring the similarity of two web pages is a complex task that requires considering multiple dimensions.
While a raw web page is an HTML document sent by the web server, web pages include \emph{Cascading Style Sheets} (CSS) that describe how the page must be rendered in the browser, and \emph{JavaScript} (JS) programs that the browser relies on to interact with the user and inject dynamic behavior into the web page.
Therefore, an effective comparison of two web pages requires exploring multiple dimensions of similarities to better detect the occurrences of any difference. 
To that end, we introduce a similarity radar that relies on the following dimensions: 1) the \emph{HTML markup} which represents the structure of the page with only the nodes of the DOM tree, 2) the \emph{HTML content} which contains all the content of the nodes of the DOM tree, 3) the \emph{JavaScript} code present in the page, 4) the \emph{CSS} code included in the page, and 5) the \emph{visual rendering} or visual similarity between two pages.
Separating a page along these dimensions ensures the comparison can be articulated around meaningful and logical parts of a page. 
This helps us pinpoint more easily the source of a difference and it also facilitates the comparison, as each dimension can have its own comparison algorithm since JS code behaves differently from CSS code and regular textual content.
Finally, visual rendering was added to understand if providing a UA header with only the string {\tt "None"}, which contains no specific device information would break a website or not.

\begin{figure}[htbp]
  \centering
  \includegraphics[width=1\linewidth]{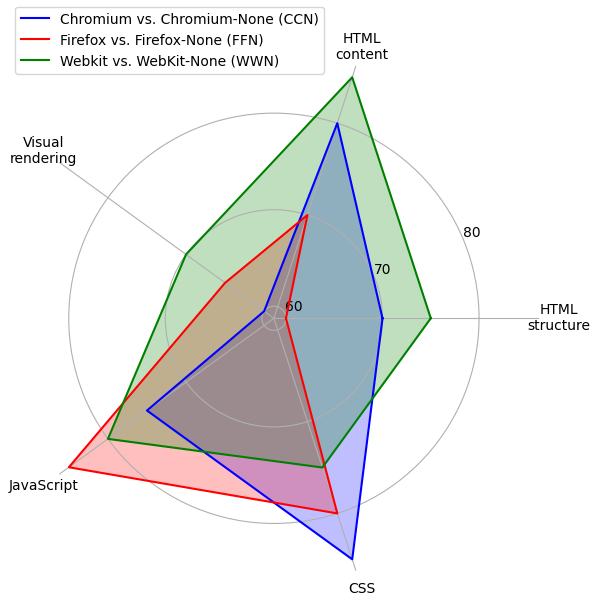}
  \caption{Similarity radar for a web page: the above represents the similarity between standard browsers and their None counterparts when accessing the home page of {\tt www.academiabarilla.it}. Each colored pentagon corresponds to a single comparison, and its vertices represent the similarity scores across five dimensions: HTML structure, HTML content, visual rendering, JavaScript, and CSS. Overlapping pentagons near the 100\% mark indicate a marginal impact of the UA on the web page.}
  \label{fig:similarity_radar_fig}
\end{figure}

In Figure \ref{fig:similarity_radar_fig}, we present the similarity radar with three colored pentagons, each representing a comparison between a standard browser and its corresponding None browser counterpart. Specifically, we compare Chromium versus Chromium-None (CCN), Firefox versus Firefox-None (FFN), and WebKit versus WebKit-None (WWN). The vertices of each pentagon are anchored to the similarity scores for the five dimensions discussed above: HTML structure, HTML content, visual rendering, JavaScript, and CSS. The further a vertex is from the center of the chart, the higher the similarity score. As such, a pentagon that covers a larger area within the chart represents a higher degree of overall similarity. In the scenario where all three pentagons overlap towards the 100\% mark on all five dimensions, the radar charts reveal that the None-browsers are highly similar to their standard counterparts. This indicates that the UA has a marginal impact on the web page. On the other hand, if there are deviations from the overlap scenario such as the FFN where the pentagon is smaller than the others, it suggests that Firefox is impacted when the UA is not known.

\begin{figure*}[htbp]
    \centering
    \includegraphics[width=1\linewidth]{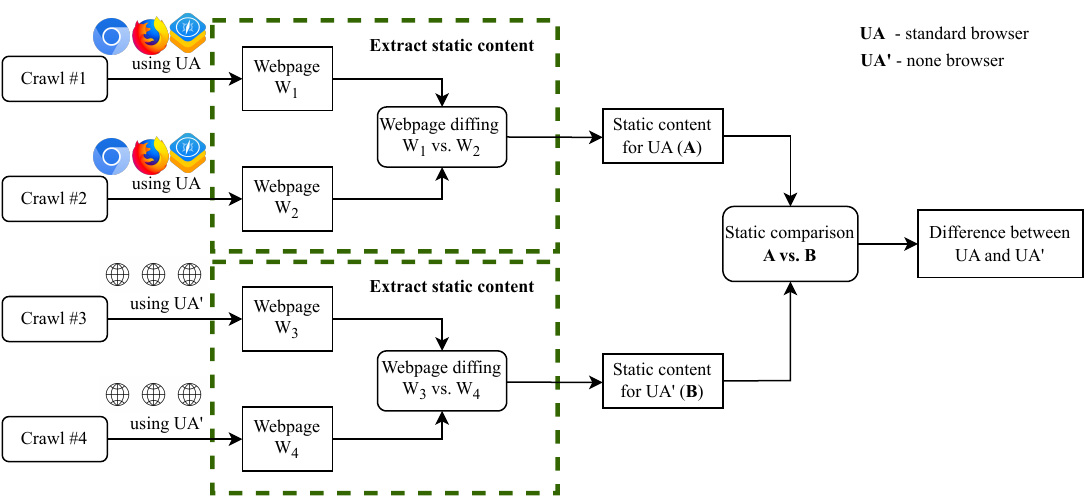}
    \caption{Highlighting web page similarity: standard browser (${UA}$) versus None-browser (${UA'}$). We crawl each web page twice using standard browsers (Chromium, Firefox, WebKit) and their None-browser counterparts. The dual crawl allows us to filter out dynamic content and focus on the static content of the web page, thereby eliminating potential bias in our analysis.  Subsequently, we execute a static comparison between standard and None-browsers' pages to identify UA-attributable differences, thereby facilitating the computation of similarity scores.}
    \label{change_analysis_fig}
\end{figure*}

\paragraph{\textbf{Removing dynamic content}}
To understand the impact of different UA, it is important to filter out the content that may differ because a page may have been visited at a different time of the day (like different articles shown on a news website) or that has been personalized based on user preferences.
To achieve this goal, we extract a backbone for each visited web page by comparing the page with itself collected from two distinct crawls.
This way, the dynamic content is revealed and can be excluded from our similarity analysis.
For example, if we visit a web page twice using a standard browser (${UA}$) and download its resources as ${W_1}$ and ${W_2}$, ${A}$ is the backbone of the visited web page such that every resource in ${A}$ belongs to both ${W_1}$ and ${W_2}$.
% \begin{displaymath}
%     A = \bigl\{ x~ |~ x \in W_1~ and~ x \in W_2 \bigr\}~ where~ A \subset W_1~ and~ A \subset W_2
% \end{displaymath}
Similarly, when investigating the impact of ${UA}$, the same page is visited twice with a None-browser (${UA'}$), and resources are downloaded as ${W_3}$ and ${W_4}$. 
${B}$ is the backbone for the web page visited using ${UA'}$, such that every resource in ${B}$ belongs to both ${W_3}$ and ${W_4}$. 
% \begin{displaymath}
%     B = \bigl\{ x~ |~ x \in W_3~ and~ x \in W_4 \bigr\}~ where~ B \subset W_3~ and~ B \subset W_4
% \end{displaymath}
Finally, a comparison between A and B results in a similarity radar for the web page visited using ${UA}$ and ${UA'}$ (cf. Fig.~\ref{change_analysis_fig}).

\subsection{Implementation Details}
Our methodology for measuring web page similarity is a combination of both existing tools and research to evaluate the similarity of the HTML structure, HTML content, JavaScript, and CSS. 
To measure the visual similarity of web pages, we propose a novel approach for comparing web page screenshots using traditional image processing techniques. 
This comprehensive approach provides an in-depth understanding of web page similarity in the wild.

\paragraph{\textbf{Removing dynamic content}} 
Every browser uses \emph{document object model} (DOM) to hierarchically organize nodes of the HTML document as a tree that renders in the browser. 
This allows structural comparison of $2$ DOM trees to capture the similarity of the HTML structure of $2$ web pages. 
Previous works have studied document structural similarity algorithms based on tree edit distance, tree matching, and various approximation techniques~\cite{bille_survey_2005, reis_automatic_2004, buttler_short_nodate}.
We use SFTM, a DOM tree matching tool, to measure the similarity of the HTML structure of two web pages in the wild~\cite{brisset_sftm}.
We choose SFTM because, in contrast to other tree-matching algorithms, the algorithm behind it efficiently processes any valid DOM tree in microseconds, not dozens of seconds.
To assess the similarity of the HTML structure of $2$ web pages, we extract the DOM trees from the HTML documents and feed them to SFTM. 
SFTM then builds a graph between the two DOM trees with edges representing edit operations on the nodes. 
We then create labels for the insertion, deletion, and replacement operations on the nodes in the graph. 
The output of the process is an object containing the number of edges (${|E|}$) and the number of edit operations (${|C|}$).

A similarity score (${S_1}$) is then computed as ${S_1 = |C| / |E|}$.

\paragraph{\textbf{Comparing the HTML content}} 
We use the Diff Match Patch library~\cite{diff_match_path} to compare the similarity of HTML content over other text comparison tools for several reasons.
Firstly, it has a high level of accuracy in detecting similarities between text snippets even in the presence of minor variations, such as white space or case sensitivity.
Additionally, the library has the capability to handle long text sequences efficiently, making it ideal for comparing HTML content which can often be lengthy.
Finally, the library offers a flexible API, allowing for customization of the comparison process to meet specific requirements, such as ignoring HTML markups in this context. 
These features make Diff Match Patch an ideal choice for evaluating the similarity of HTML content. 
The Diff Match Patch library implements Myer's diff algorithm~\cite{myers_anond_1986}. 
To assess the similarity of the HTML content of $2$ web pages, we feed the raw content of the HTML documents to Diff Match Patch. 
The library returns a graph of edges (${E}$) with edit operations on the nodes. 
Diff Match Patch uses the Levenshtein distance (${d}$) to compute the number of edit operations between $2$ streams of characters from the HTML documents. 
A similarity score (${S_2}$) is then computed such as ${S_2 = d / |E|}$~\cite{yujian_normalized_2007}.

\paragraph{\textbf{Comparing JS \& CSS}} 
To assess the similarity of JS or CSS on $2$ web pages, we need to first build an \emph{abstract syntax tree} (AST) from each JS script and CSS stylesheet. 
We apply a \emph{Locality-Sensitive Hash} (LSH) function on the content of each file to determine specific files with changes~\cite{datar_locality-sensitive_2004}. 
Once we extract ASTs to compare, similarly to the process of comparing the HTML structure, we conduct a tree-matching process to obtain a graph of edges and edit operations on nodes of the AST trees. 
We use \textsc{GumTree}, an AST diff tool that allows a plug-and-play of language parsers, to compare the AST trees~\cite{falleri_gumtree}. 
We then create labels for the insertion, deletion, and replacement operations on the nodes in the resulting graph. 
The output of the process is an object containing the number of edges (${|E|}$) and the number of edit operations (${|D|}$) reported by \textsc{GumTree}.
%and the number of actions done with the edit operations (${A}$). 
A similarity score (${S_3}$) is then computed such as ${S_3 = |D| / |E|}$.

\paragraph{\textbf{Comparing visual rendering}} 
State-of-the-art image comparison algorithms are based on perceptual hashing, histogram, or by looking at pixel-by-pixel changes~\cite{drmic_evaluating_2017, kotsarenko_measuring_2018, song_multiple_2011, vallez_needle_2022, vysniauskas_anti-aliased_2009}.
While those algorithms can detect the smallest difference when comparing pictures or screenshots of web pages, they fail to capture more macro changes, like text changes, broken links, or missing images. 
We introduce a novel approach to compare the visual rendering of web pages based on the Canny edge detection algorithm to detect any object or shape in the screenshot, which can represent text, multimedia content, and visual sections of the web page regardless of its size~\cite{canny_computational_1986}. 
Our algorithm computes the number of edges (contours) in a screenshot of the web pages, hence calling it contour-based analysis.

\begin{figure}
    \centering
    \subfloat{{\includegraphics[width=.89\linewidth]{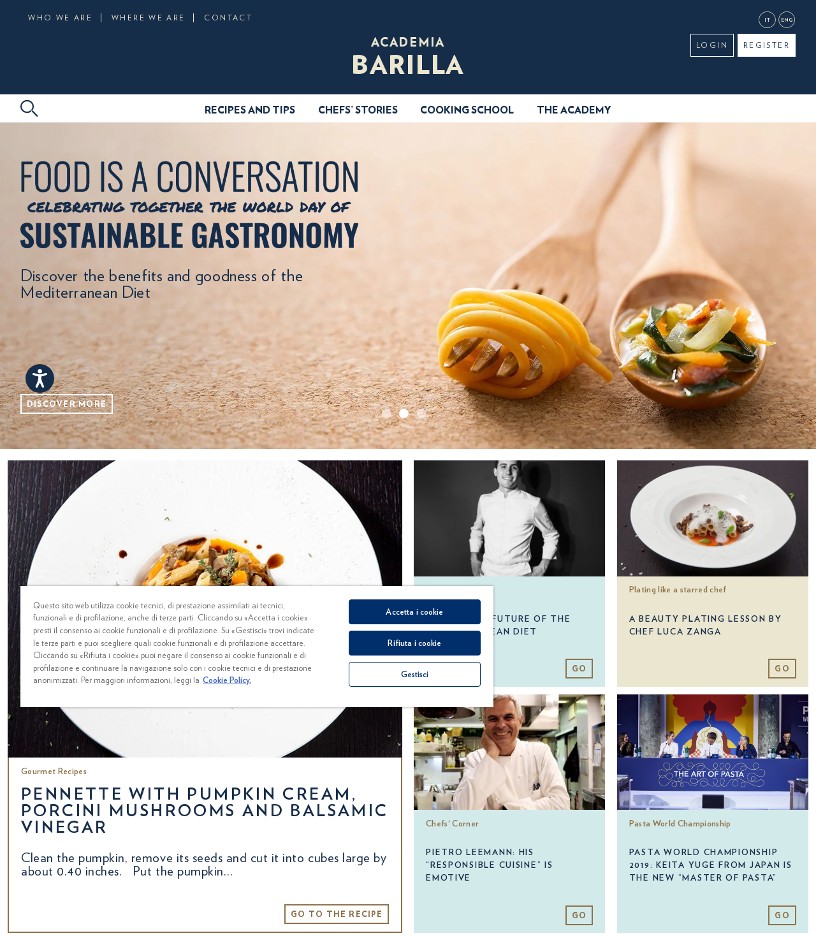} }}
    \qquad
    \subfloat{{\includegraphics[width=.89\linewidth]{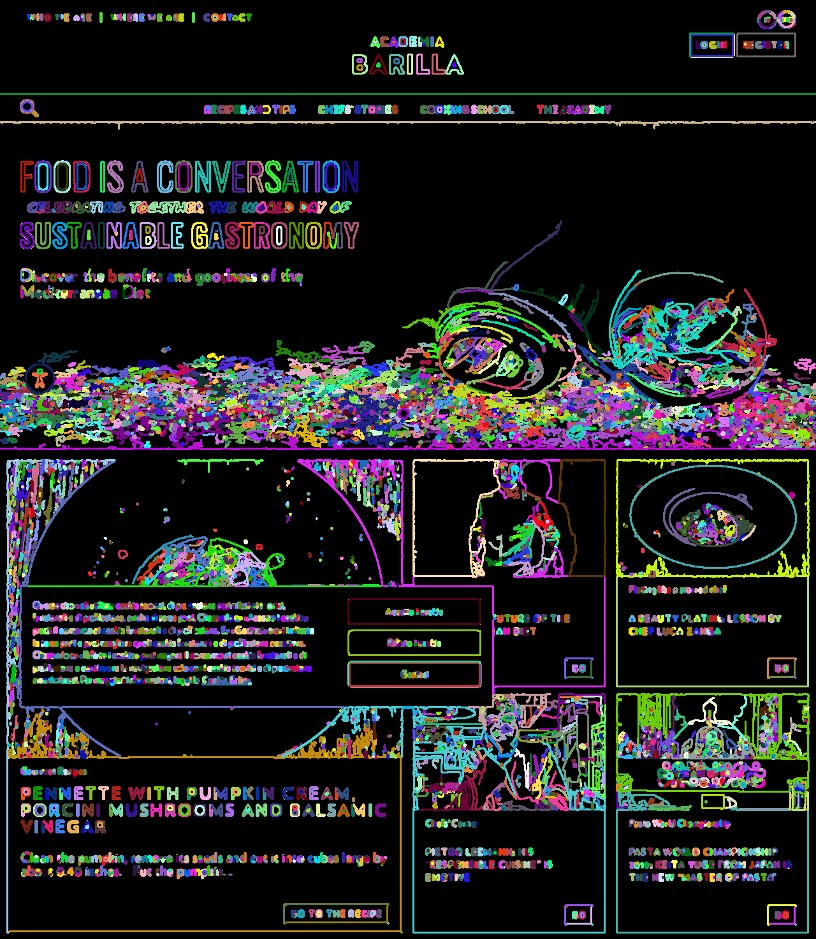} }}
    \caption{Contour-based visual analysis: the figure illustrates the process of contour-based analysis on a screenshot taken from {\tt www.academiabarilla.it}. The top image represents the original screenshot, while the bottom image shows the identified contours (edges), representing different objects and shapes within the web page. This technique enables a comparison of visual rendering, capturing significant changes such as text modifications, broken links, or missing images.}
    \label{contours_analysis_fig}
\end{figure}

We rely on OpenCV~\cite{opencv_website} to retrieve contours from an image. 
OpenCV implements the shape analysis algorithm by Satoshi~\emph{et~al.}~\cite{suzuki_topological_1985}.
For better accuracy, we convert the original screenshot into a binary image and then find the contours in the image.
Using the contours in the image, we compute the areas of the contours and their \emph{moments}.
In OpenCV, moments are the average of the intensities of an image’s pixels.
The area of a contour gives it relevance compared to other contours in the image while the moment helps us determine the difference in the same contour.
For example, web pages on news websites often have the same contours, but with different text and photos in the same placeholder.
Using moments helps us determine if the content within the same contour has changed. 
Algorithm~\ref{find_contours_algo} shows the steps we take to compute the properties of our image contour analysis, where \emph{s} is the file path of the screenshot, while \emph{cv::cvtColor}, \emph{cv::findContours}, \emph{cv::contourArea}, \emph{cv::boundingRect} are arrays computed using OpenCV functions.
The image contour properties that we compute are the number of contours (${|C|}$), the weighted aggregate of contour areas (${A}$), and the weighted aggregate contour moments (${M}$).

\begin{algorithm}
    \caption{Contour properties: this algorithm takes an input image $s$, converts it to a grayscale image $g$, finds contours $C$ from the grayscale image, stores the area of each contour in $Y$, calculates the bounding rectangle $Z$, and computes the weighted areas $A$ and moments $M$ of the contours.}
    \label{find_contours_algo}
    \begin{algorithmic}[1]
      \Function{FindContourProperties}{$s$}
        \State ${g} = \text{cv::cvtColor}(s)$
        \Comment{Convert image $s$ to grayscale image $g$}
        \State ${C} = \text{cv::findContours}(g)$
        \Comment{Find contours $C$ from image $g$}
        \State ${Y} = \text{cv::contourArea}(C)$
        \Comment{Store area of each contour in $Y$}
        \State ${Z} = \text{cv::boundingRect}(C)$
        \Comment{Bounding rectangle $Z$}
        \State Let $A = 0$, $M = 0$
        \Comment{Weighted areas $A$ \& moments $M$}
        % \State Let $n$ be the size of $X$
        % \For{$i = 1$ to ${|X|}$}
        %     \State $C = C + X_i$
        % \EndFor
        \For{$i = 1$ to ${|C|}$}
            \State $A = A + {Y_i^2} \div Y$
            \State $M = M + {Z_i^2} \div Z$
        \EndFor
        \State \Return <|C|, A, M>
       \EndFunction
    \end{algorithmic}
\end{algorithm}

Since the contour properties (${C}$, ${A}$, and ${M}$) are heterogeneous, when comparing two screenshots to find their similarity, we compute the geometric mean (${GM}$) of the contour properties of each screenshot.
Finally, we compute the visual similarity score (${S}$) as the ratio of the absolute difference of the geometric means (${GM_1}$ and ${GM_2}$) to their arithmetic mean such that:

  \begin{equation}
    GM = \sqrt[\leftroot{0}\uproot{2}3]{|C| \times A \times M}
  \end{equation}
  
  \begin{equation}
    S = \dfrac{|GM_1 - GM_2|}{(GM_1 + GM_2)/2}
  \end{equation}

\section{Exploring the Impact of UA Changes}\label{dataset}
To explore the impact of UA on the web, we crawled websites with the default HTTP's UA header and with the {\tt "None"} string in its place. 
We conducted regression tests based on the similarity radars provided by \radar. 
We then analyzed the edit operations for the dimensions in each test to determine if the observed changes were due to the removal of identifying information in the UA or not. 
We also explore in this section why those changes occurred.

\subsection{Crawl Description \& Statistics}
We used a web testing and automation framework called {\em Playwright} to instrument standard browsers, namely Chromium (${C}$), Firefox (${F}$), and Safari (with the WebKit engine ${W}$)~\cite{playwright_website}. 
To instrument the None browsers, we modified the HTTP request-header field {\tt User-Agent} of the standard browsers and changed it to the word \texttt{"None"}. 
We also modified information that identifies the browser in the Navigator objects of the standard browsers. 
In particular, we changed \texttt{navigator.appVersion}, \texttt{navigator.platform}, \texttt{navigator.userAgent}, and\\ \texttt{navigator.vendor} and placed the word \texttt{"None"} on each of those properties. 
Furthermore, to avoid our modified browsers from being detected as bots, we set \texttt{navigator.webdriver} to \texttt{false}. 
Detailed lists of navigator properties exposed on all browsers during the crawl are available in the artifacts mentioned in Section~\ref{appendices}. 
We called the resulting browsers after their modified versions: \emph{Chromium-None} (${CN}$), \emph{Firefox-None} (${FN}$), and \emph{WebKit-None} (${WN}$). 
In the end, we ran the crawl with $6$ browsers in total.

After preparing the browsers to be used, we decided on how to run the crawl. 
We used the \textsc{Tranco} list to choose the domains to crawl~\cite{tranco_website}. 
We chose the \textsc{Tranco} list as its ranking of website popularity surpasses other sources of web traffic analysis~\cite{pochat_tranco_2019}. 
Nevertheless, previous studies have expressed concerns about the methodology used to create popular lists, such as \textsc{Tranco} and their representativeness~\cite{scheitle_long_2018}. 
For that reason, we randomized our crawl of domains on the \textsc{Tranco} list until we reached the limit of our computing resources. 
We finally crawled homepages of $12,000$ domains with $1,765$ in the Top\,10k domains, $6,036$ between the Top\,10k and Top\,100k, and $4,199$ between the Top\,100k and Top\,1M.
Aqeel~\emph{et~al.} have also questioned the representativeness of measurement studies that rely only on landing pages and no internal pages, citing a difference in structure and content between the landing page and the internal pages\cite{aqeel_landing_2020}.
This was not a concern for our study as our objective was to analyze the impact of restricting the UA without being specific on the type of structure or content.

We crawled the homepage of each domain four times with each browser: twice (for self-comparison to remove dynamic content) before the execution of JS to study differential serving, and twice after the execution of JS to study content adaptation. 
This represents $24$ visits per homepage in total. 
For differential serving, we waited for the ${domcontentloaded}$ event to be fired before saving on disk the complete HTML document with all first and third-party JS and CSS files. 
For content adaptation, we waited 15 seconds after the ${load}$ event was fired to save everything on disk.
To avoid being rejected due to a high number of requests, our crawler sent exactly one request to one domain with one browser at a time, and only multiple requests to multiple domains in parallel. 
The crawl ran for $1$ month. 
A repository for the dataset of this paper is available in the artifacts mentioned in Section~\ref{appendices}.

% Due to storage constraints, we compressed all the downloaded files using Bzip2, an open-source tool that implements a block-sorting compression algorithm~\cite{noauthor_bzip2_nodate}. 
% We chose Bzip2 not only because it is open source but also because of its high compression ratio, particularly for compressing large files. 
% Additionally, Bzip2 is designed for fast decompression, which was important in our case as we needed to decompress files for the \radar comparison. 
If all  $24$ requests were not successfully crawled with the data correctly saved, the crawled domain was ignored and all downloaded resources for the crawled web page were deleted on the disk. 
In the end, we successfully crawled and saved data for $270,048$ web pages from $11,252$ domains. 
We stored $5.85$ Terabytes of compressed files on the disk. 
Table~\ref{crawled_resources} summarizes the resources we saved on the disk during our crawl. 
It is worth noting that JS takes most of the resources on the Internet today with 73\% of the downloaded files and 80\% of disk space in our dataset.

\begin{table}[!ht]
    \centering
    \caption{Summary of crawled resources}
    \label{crawled_resources}
    \begin{tabular}{l|r|r}
    \hline
        \bf Resource type   & \bf Number of files   & \bf Resource size \\ \hline
        HTML                & $180,032$             & 17 GB \\
        JavaScript          & $73,573,872$          & $4,705$ GB \\
        CSS                 & $27,060,120$          & 959 GB \\
        Screenshots         & $180,032$             & 167 GB \\ \hline \hline
        \bf Total:          & $\bf 100,994,056$     & $\bf 5,848$ \bf GB
    \end{tabular}
\end{table}

\subsection{Empirical Results \& Findings}
In this section, we use the following notations: \textsf{CCN} for the comparison between pages from Chromium\ against \,Chromium-None, \textsf{FFN} for Firefox\ against \,Firefox-None\ and \textsf{WWN} for WebKit\, against \,WebKit-None.\\

\begin{figure*}[htbp]
    \centering
    \subfloat[All\label{fig:all_radar}]{{\includegraphics[width=5.6cm]{{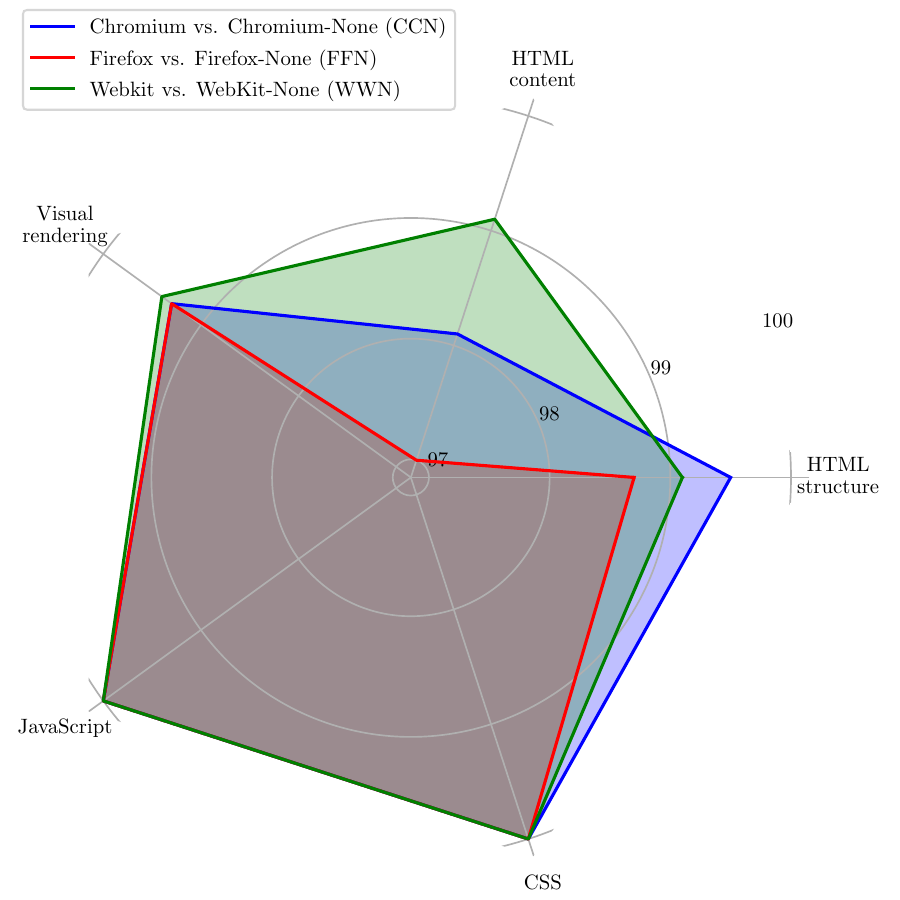}}}}
    \hfill
    \subfloat[News\label{fig:news_radar}]{{\includegraphics[width=5.6cm]{{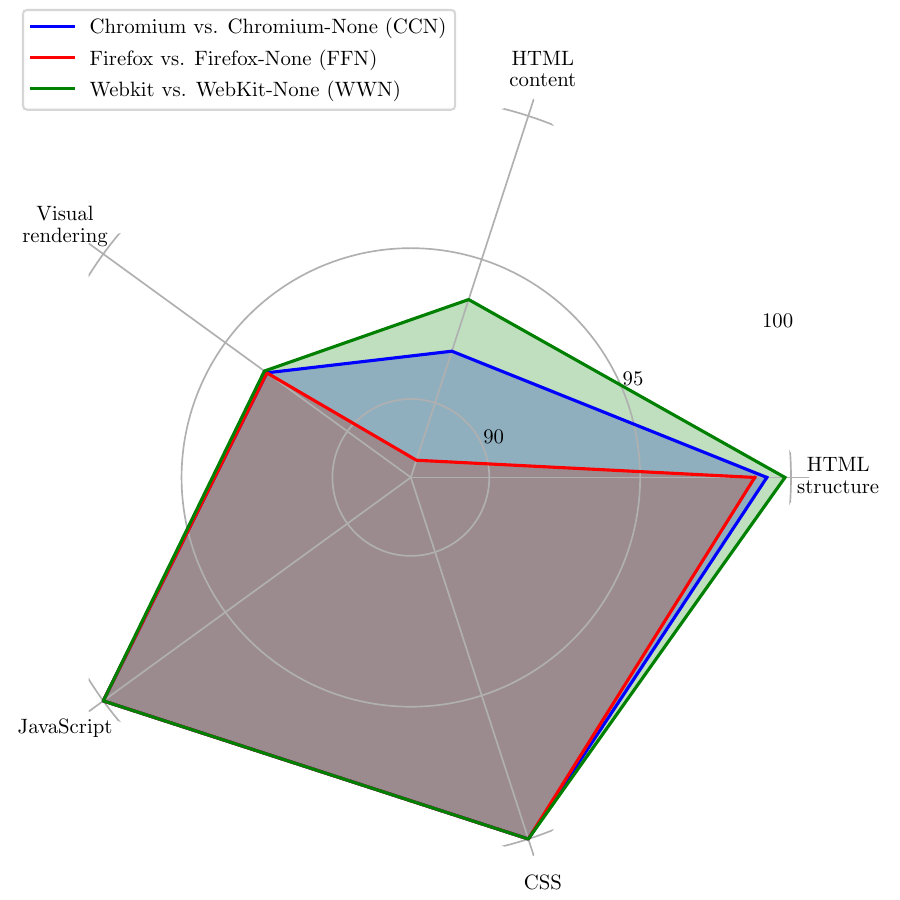}}}}
    \hfill
    \subfloat[Internet Services\label{fig:internet_radar}]{{\includegraphics[width=5.6cm]{{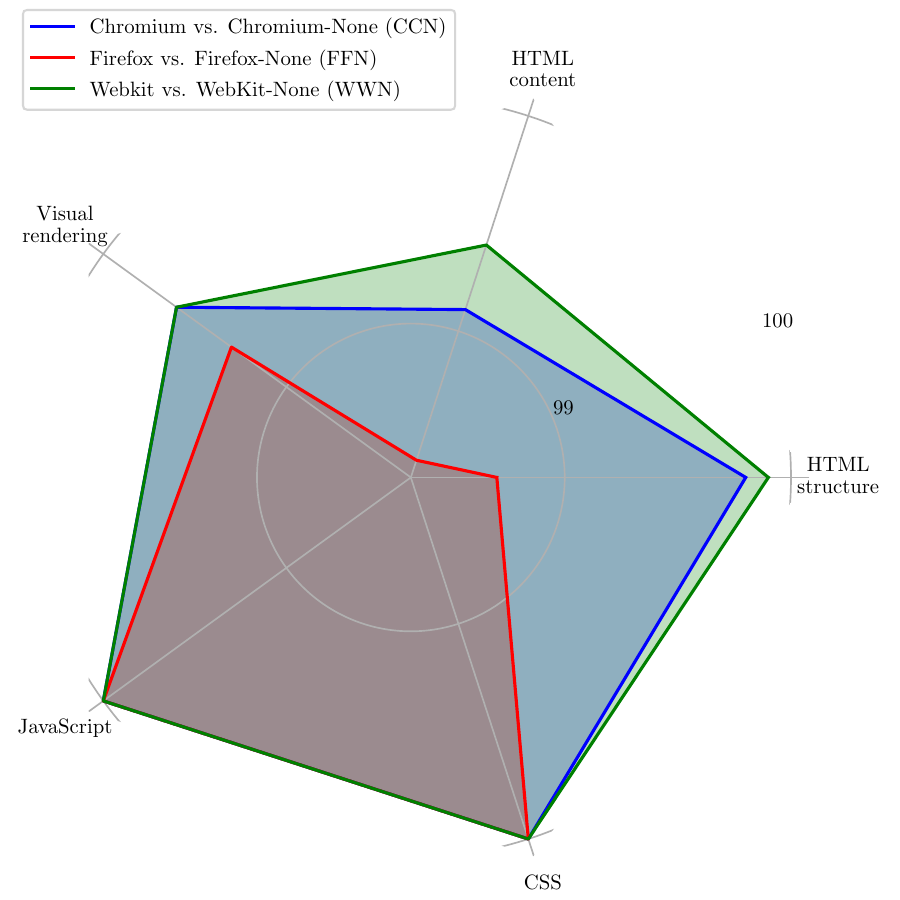}}}}
    \\
    \subfloat[Business\label{fig:business_radar}]{{\includegraphics[width=5.6cm]{{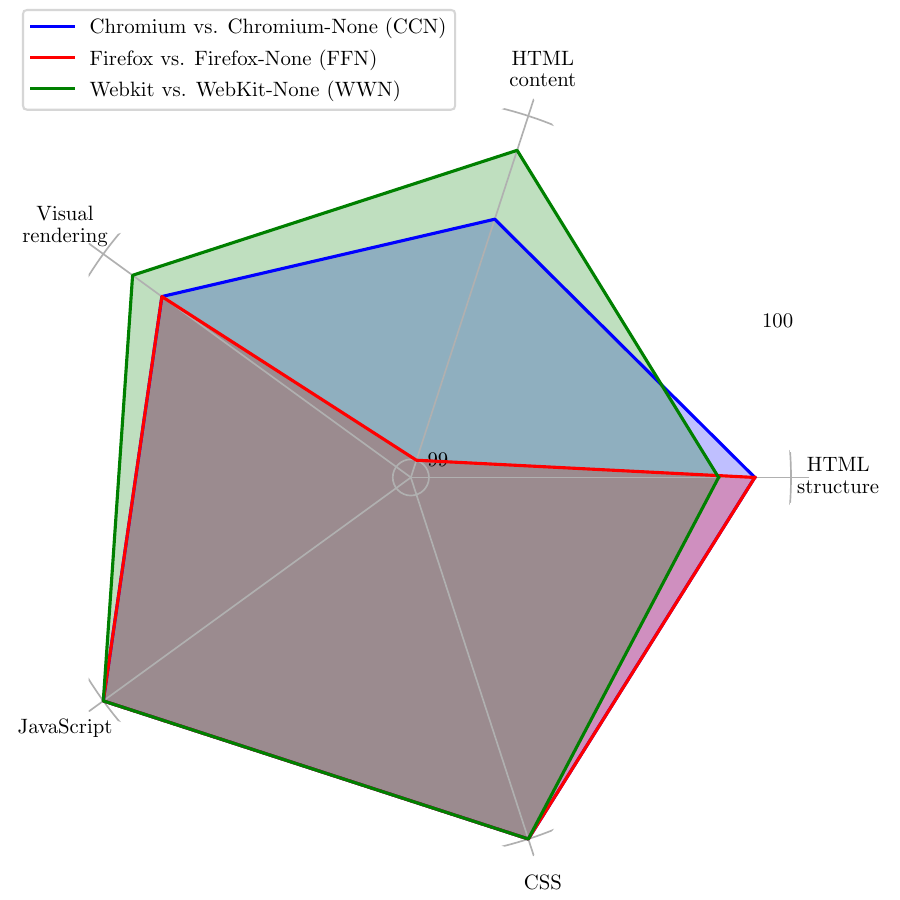}}}}
    \hfill
    \subfloat[Marketing\label{fig:marketing_radar}]{{\includegraphics[width=5.6cm]{{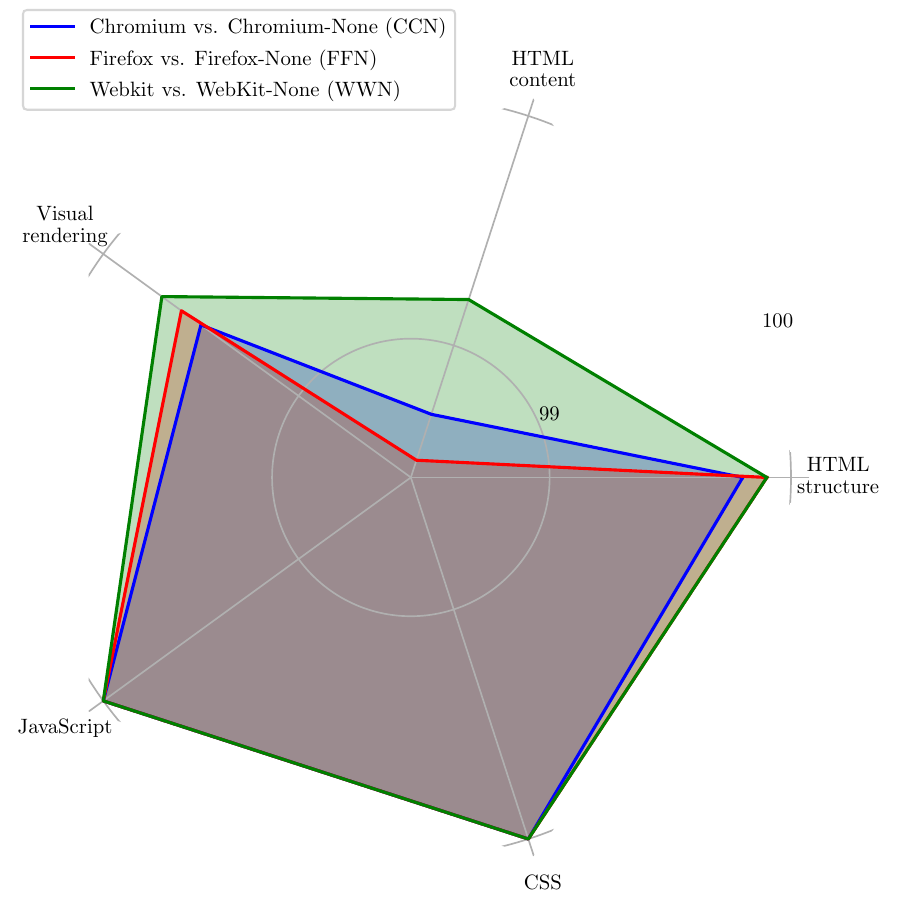}}}}
    \hfill
    \subfloat[Online Shopping\label{fig:shopping_radar}]{{\includegraphics[width=5.6cm]{{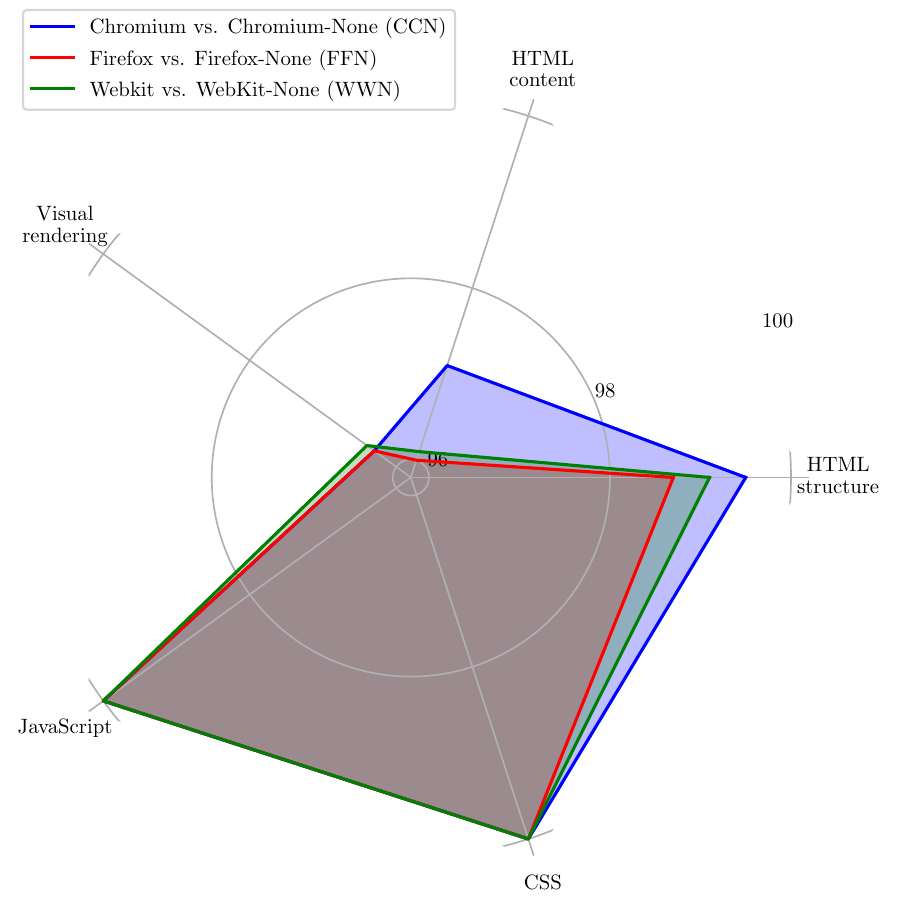}}}}
    \caption{Average similarity scores across website categories: this figure illustrates the average similarity scores between standard browsers and their None-browser counterparts for the top five categories in our dataset.}
    \label{fig:similarity_scores}
\end{figure*}

\paragraph{\textbf{Differential serving}}
The average similarity scores before JS is executed are 100\% for all the tested browsers (${CCN}$, ${FFN}$, and ${WWN}$) on HTML structure, HTML content, JS, and CSS (cf. Figure~\ref{fig:all_radar}). 
One takeaway is that web servers reply to all HTTP requests with the same HTML document, regardless of the fact that the UA in the HTTP request header is known or not. 
This is possible because browsers adopt the same standards, such as responsive web design to adjust the rendering of web pages to browsing environments. 
That means that, nowadays, websites focus on consistent user experience across devices and browsers rather than device-specific content. 
The fact that UA is no longer the sole factor in determining the content served makes them less relevant and hence removing the significance of the UA can reduce the attack surface and improve the privacy and security of users.

\paragraph{\textbf{Content adaptation}} 
When JavaScript is executed, the average similarity scores remain 100\% on JS and CSS. 
However, there are changes in visual rendering, the HTML structure, and the HTML content. 
$158$ out of the $11,252$ domains are not 100\% visually similar for at least one of ${CCN}$, ${FFN}$, or ${WWN}$. 
Going with the same logic, $955$ are not 100\% similar on both the HTML structure and HTML content for at least one of ${CCN}$, ${FFN}$, ${WWN}$. 
We looked up the $158$ domains with at least one difference and found that all those domains also have at least one difference on both HTML structure and HTML content for at least one of ${CCN}$, ${FFN}$, or ${WWN}$. 
We then established that, out of $11,252$ domains, $955$ are changed by the lack of a known UA or other identifying information. 
That is, 8.4\% of our dataset was dependent on the UA.
Table~\ref{ua_dependent_categories_table} lists the Internet categories to which the UA-dependent domains belong.
We used McAfee SmartFilter to obtain the Internet categories of the UA-dependent domains~\cite{mcafee_url_categorization}.

\begin{table}[h!]
  \centering
      \caption{Internet categories of the UA-dependent domains}\label{ua_dependent_categories_table}
      \begin{tabular}{lr}
      \hline
          \bf Category & \multicolumn{1}{r}{\begin{tabular}[c]{@{}r@{}}\bf Number of\\\bf Domains\end{tabular}} \\ \hline
          News & 180 \\
          Internet Services & 156 \\
          Business & 128 \\
          Online Shopping & 109 \\
          Marketing & 106 \\
          Blogs & 89 \\
          Education & 71 \\
          Entertainment & 68 \\
          Information & 38 \\
          Finance & 10 \\ \hline \hline
          \bf Total & \bf 955 \\
      \end{tabular}
\end{table}

\begin{table}[h!]
      \centering
      \caption{Summary of problem severity levels for UA-dependent domains}\label{problem_severity_table}
      \begin{tabular}{l|r}
      \hline
          \bf Category & \bf Number of domains \\ \hline
          IRRITANT & 225 \\
          MODERATE & 131 \\
          NO PATTERN & 6 \\
          SEVERE & 526 \\
          UNUSABLE & 67 \\ \hline \hline
          \bf TOTAL & \bf 955 \\
      \end{tabular}
\end{table}

The last $5$ radars depicted in Figure~\ref{fig:similarity_scores} provide a category-wise breakdown of the average similarity scores between standard browsers and their None-browser counterparts for the top\,5 website categories in our data set, namely: news, internet services, business, online shopping, and marketing.
In the category "news" (cf. Figure~\ref{fig:news_radar}), the HTML content and visual rendering dimensions have lower similarity scores compared to other dimensions.
This suggests that, in this category, the absence of a known UA tends to influence the visual presentation and HTML content of the web pages more significantly.
For the "internet services", "business", and "marketing" categories  (cf. Figures~\ref{fig:internet_radar},~\ref{fig:business_radar}, and~\ref{fig:marketing_radar}), all three comparisons (CCN, FFN, and WWN) show similarity scores gravitating towards 100\% across all dimensions. 
This indicates the similarity of HTML content and visual rendering, irrespective of the browser's UA.
Therefore, we can infer that the impact of the UA on these categories is marginal.
The category "online shopping" (cf. Figure~\ref{fig:shopping_radar}), however, presents a contrast.
The HTML content and visual rendering dimensions have lower similarity scores, close to the category "news", but we observe a departure from the pattern observed in other categories. 
While, the similarity scores for the WebKit browser usually top the charts, followed by Chromium, in the category "online shopping" the HTML content similarity score for WebKit drops dramatically, coming closer to Firefox.
Additionally, the visual rendering is impacted across all browsers. This may suggest that online shopping websites employ more complex or diverse techniques for content adaptation based on UA, which could potentially be linked to the need for enhanced user experience or functionality specific to the website's purpose.

The consistent performance of WebKit, then Chromium across categories prompts further exploration. One plausible explanation could lie in the rendering engine used by the browsers. Both browsers use Blink, however, WebKit consistently outperforms Chromium in our analysis. While this difference could stem from how each browser integrates and uses the Blink engine, the underlying reasons for this consistent trend are not immediately apparent from our study and would require further investigation. Such research could provide insights into how browser architecture and rendering engines influence content adaptation.

\begin{algorithm}
    \caption{Change impact analysis: this algorithm assesses the impact on a web page when changes occur due to the use of a None-browser. It evaluates the differences detected in the static comparison (as illustrated in Figure \ref{change_analysis_fig}) of both standard and None-browsers. Subsequently, these differences are categorized based on specific patterns that we identified during manual analysis.}
    \center
    \label{find_change_impact_algo}
    \begin{algorithmic}[1]
      \Function{FindChangeImpact}{$C,CN,F,FN,W,WN$}
        \State Let $\Delta CCN \leftarrow \text{StaticComparison}(C, CN)$
        \State Let $\Delta FFN \leftarrow \text{StaticComparison}(F, FN)$
        \State Let $\Delta WWN \leftarrow \text{StaticComparison}(W, WN)$
        \State Let $\Delta CF \leftarrow \text{StaticComparison}(C, F)$
        \State Let $\Delta CW \leftarrow \text{StaticComparison}(C, W)$
        \State Let $\Delta FW \leftarrow \text{StaticComparison}(F, W)$
        \If {$\Delta CCN = \Delta FFN\ \&\ \Delta FFN = \Delta WWN$}
            \State Let $R \leftarrow [\Delta CCN, \Delta FFN, \Delta WWN]$
            \State Let $N \leftarrow [\Delta CF, \Delta CW, \Delta FW]$
            \If {margin-\{top,bottom\} $\in \Delta CCN$}
                \State $\Return \text{"Margin\ collapsing\ fail"}$
            \ElsIf {white-space: wrap $\in \Delta CCN$}
                \State $\Return \text{"Soft-wrap\ fail"}$
            \ElsIf {page-break-{before,after} $\in \Delta CCN$}
                \State $\Return \text{"Unnecessary\ blank \ lines"}$
            \ElsIf {<tag\ css> $\in \Delta CCN$}
                \State $\Return \text{"Inline\ css \ changes"}$
            \ElsIf {<img\ src> $\in \Delta CCN$}
                \State $\Return \text{"Lazy\ loading\ fail"}$
            \ElsIf {<iframe\ width|height> $\in \Delta CCN$}
                \State $\Return \text{"Displaced\ iframe"}$
            \ElsIf {<tag\ inactive|disabled> $\in \Delta CCN$}
                \State $\Return \text{"Disabled\ component"}$
            \ElsIf {CAPTCHA|403|error $\in \Delta CCN$}
                \State $\Return \text{"Browser\ not\ identified"}$
            \ElsIf {$R \neq N$}
                \State $\Return \text{"Content\ restriction"}$
            \Else
                \State $\Return \text{"No\ pattern"}$
            \EndIf
        \Else
        \State Let $x \leftarrow \Delta CCN \neq \Delta FFN \neq \Delta WWN$
        \State Let $y \leftarrow \Delta CF \neq \Delta CW \neq \Delta FW$
            \If {$x$ \& $y$}
                \State $\Return \text{"No\ impact"}$
            \Else
                \State $\Return \text{"No\ pattern"}$
            \EndIf
        \EndIf
        % \State \Return i
      \EndFunction
    \end{algorithmic}
\end{algorithm}

\paragraph{\textbf{Changes created by different UA and their causes}} 
Due to the intensive manual efforts required to analyze the changes, out of the $955$ domains that changed because of the None-browser, we manually analyzed the changes for $204$ domains. 
We detected 10 patterns of the impact of those changes and used the 10 patterns to apply heuristics to the rest of the data set in order to classify the severity of the impact on the usability of the web pages. 

To build that classification, we borrowed the taxonomy of problem severity scale in usability by Rubin~\emph{et~al.}~\cite{rubin_handbook_2008}. 
Algorithm~\ref{find_change_impact_algo} details the steps in conducting the heuristics for that classification. The algorithm takes six inputs: \(C\), \(CN\), \(F\), \(FN\), \(W\), and \(WN\), which represent the used browsers: Chromium, Chromium-None, Firefox, Firefox-None, WebKit, and WebKit-None, respectively. It then performs static comparison operations between each standard browser and its None counterpart to determine the differences, denoted as \(\Delta CCN\), \(\Delta FFN\), and \(\Delta WWN\).

\begin{figure*}[htbp]
    \centering
    \subfloat[Standard browser - normal margin rendering]{{\includegraphics[width=8.5cm]{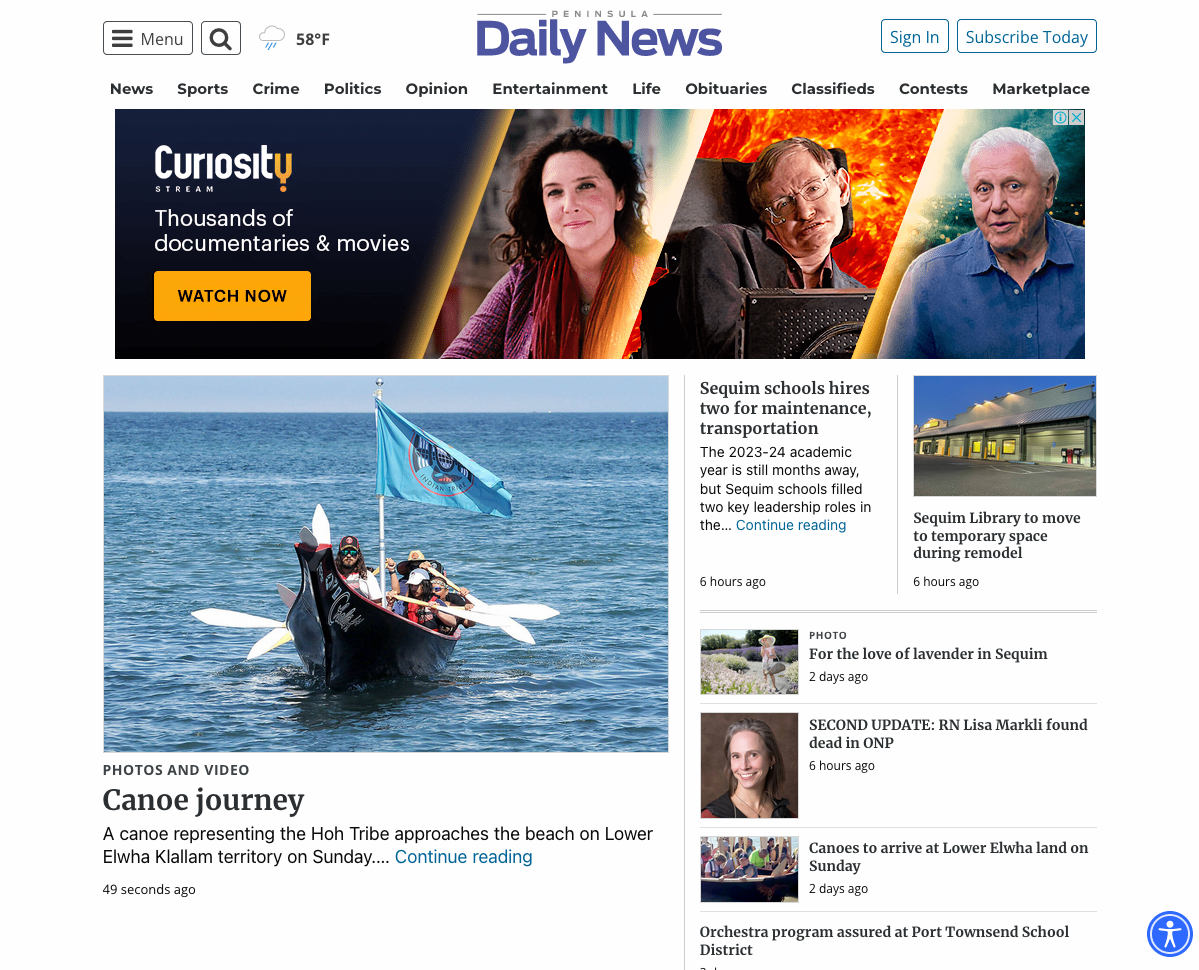}}}
    \hfill
    \subfloat[None browser - failure of margin collapse]{{\includegraphics[width=8.5cm]{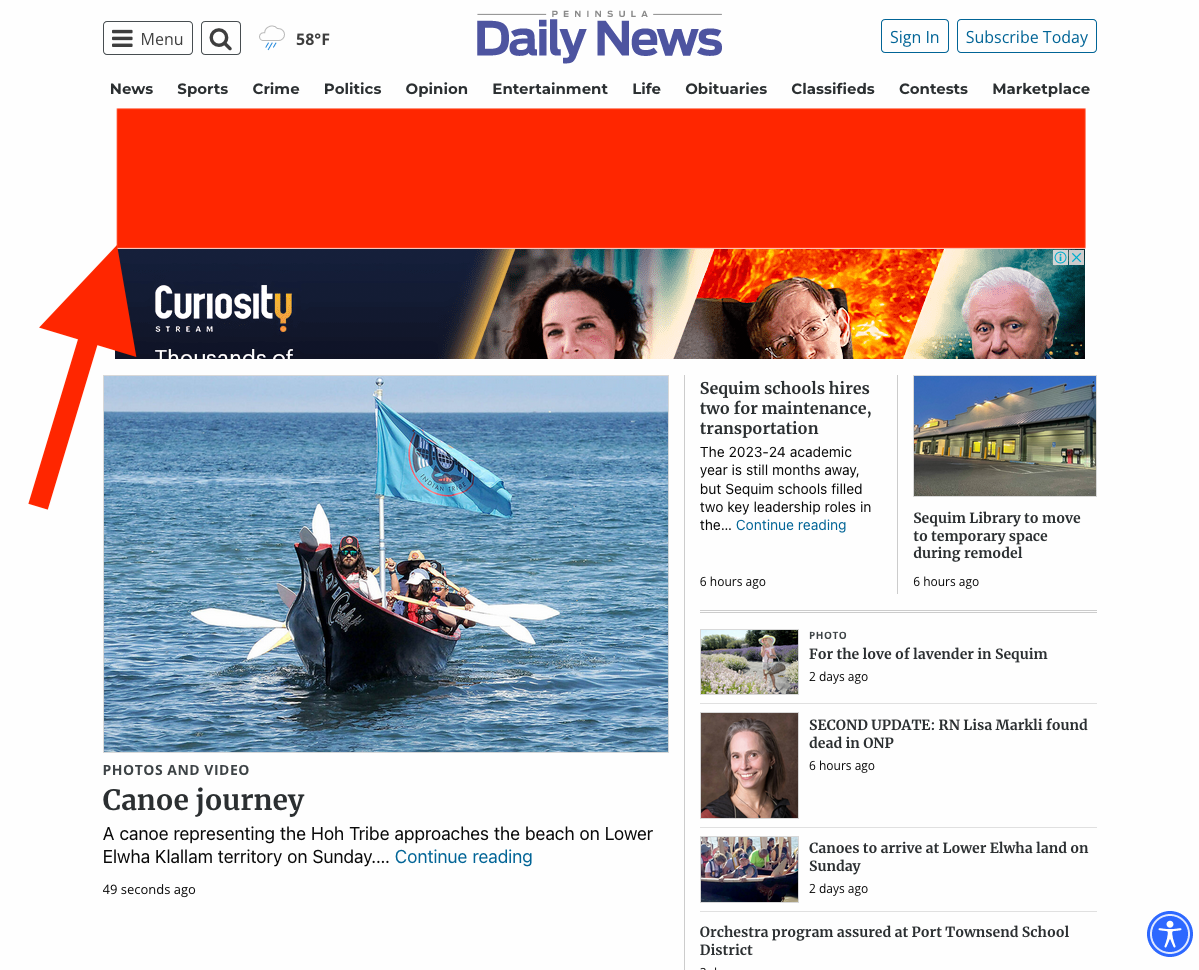}}}
    \caption{Comparison of web page rendering with standard and none browsers, illustrating a 'severe' problem severity case where a failure of margin collapse occurs in the none browser. The affected area is highlighted in red.}
    \label{fig:severe_cases}
\end{figure*}

If all three differences are identical, the algorithm then computes the differences between every pair of standard browsers, denoted as \(\Delta CF\), \(\Delta CW\), and \(\Delta FW\). It also assigns the three differences from the standard to None comparisons to the list \(R\), and the differences from the standard to standard comparisons to the list \(N\). Following this, the algorithm checks for specific CSS and HTML properties and attributes in the differences. These include the CSS properties "margin-{top, bottom}", "white-space: wrap", "page-break-{before, after}", any CSS attributes associated with HTML tags, the image source attribute, and the width or height attributes of iframes. For each of these, if they are found in the differences, the algorithm returns a corresponding impact statement. To select the properties and attributes that the algorithm used, we conducted a manual analysis of 100 websites to build a list of HTML and CSS properties that cause changes in the web page when the UA is not known. Afterward, we utilized that list to classify the levels of problem severity identified by our change impact analysis.

\begin{lstlisting}[caption=Example of unintentional restriction: a function that relies on known UA (parameter ${b}$) in a script from Google Ad Manager., label=margin]
function rn(a, b, c, d) {
  O(a.K, {
    transition: c / 1E3 + "s",
    "transition-timing-function": d,
    "margin-top": b
  })
}
\end{lstlisting}

Additionally, if a disabled or inactive tag is found, or any instance of CAPTCHA, HTTP 403 error is detected, the algorithm will return the respective impact statements. If the differences in \(R\) are not identical to this \(N\), the algorithm returns "content restriction". If none of the specific properties or attributes is found, it returns "no pattern". If the three differences are not identical, the algorithm checks if each difference is unique and returns "no impact" if that is the case. Otherwise, it returns "no impact". 

Below are the definitions of the problem severity categories that we use:
\begin{enumerate}
\item \textsf{Irritant}: The problem occurs only intermittently, can be circumvented easily, or is dependent on a standard that is outside the product’s boundaries. Could also be a cosmetic problem.	
\item \textsf{Moderate}: The user will be able to use the product in most cases, but will have to undertake some moderate effort in getting around the problem.
\item \textsf{Severe}: The user will probably use or attempt to use the product here, but will be severely limited in his or her ability to do so.	
\item \textsf{Unusable}: The user is not able to or will not want to use a particular part of the product because of the way that the product has been designed and implemented.
\end{enumerate}

\begin{table*}[htbp]
\centering
\caption{Change impact analysis of the 955 UA-dependent websites: this table details the specific changes detected, their associated impact, the problem severity level, and the number of occurrences, providing a comprehensive overview of how changes in the UA affect different aspects of the web page.}
\label{change_impact_analysis_table}
\begin{adjustbox}{width=\linewidth}
\small
\begin{tabular}{l | l | c | r}
 \hline
 \bf Change                                                                                                  & \bf Impact                    &\bf Severity &\bf Occurences\\ \hline
 CSS property: \texttt{margin-\{top,bottom\}}                                                                & Failure of margin collapsing  & SEVERE   & 252\\
 $\{{\Delta}CCN {\ne} {\Delta}FFN {\ne} {\Delta}WWN$\} \& $\{{\Delta}CF {\ne} {\Delta}CW {\ne} {\Delta}FW\}$ & No impact                     & IRRITANT & 225\\
 Missing image \texttt{SRC} reference                                                                        & Failure of lazy loading       & SEVERE   & 101\\
 CSS property: \texttt{white-space: wrap}                                                                    & Failure of soft-wrap          & SEVERE   &  99\\
 Change of CSS attribute(s)                                                                                  & Change of inline CSS          & MODERATE &  83\\
 CSS property: \texttt{page-break-\{before,after\}}                                                          & Unnecessary blank lines       & SEVERE   &  74\\
 iFrame \texttt{width} | \texttt{height}                                                                     & Displaced iframe              & MODERATE &  48\\
 $\{{\Delta}CCN = {\Delta}FFN = {\Delta}WWN\} \ne \{{\Delta}CF = {\Delta}CW = {\Delta}FW\}$                  & Content restriction           & UNUSABLE &  38\\
 CAPTCHA or 403 Error or Browser Error                                                                       & Browser not identified        & UNUSABLE &  22\\
 CSS \texttt{:disabled} | \texttt{:inactive}                                                                 & Disabled component            & UNUSABLE &   7\\
 No pattern                                                                                                  & -                             & -        &   6\\
 \hline
\end{tabular}
\end{adjustbox}
\end{table*} 

It should be noted that we do not address the behaviour and interaction with the functionality of the web page, so the use of the word "UNUSABLE" in the problem severity scale by Rubin~\emph{et~al.} should not create confusion. 
Table~\ref{change_impact_analysis_table} shows the classes of the impact of the changes and the severity of the impact. 
On $425$ out of the $955$ domains (44.5\%), the impact of browsing those domains with a None-browser was spacing issues (failure of margin collapsing, failure of soft-wrap, and unnecessary blank lines), while on $515$ domains, the impact was driven by CSS issues. 
Table~\ref{problem_severity_table} shows the distribution of problem severity for the $955$ domains, revealing that just 7\% of the domains were unusable when we browsed them using a None-browser.

Some HTML elements change while the browser adjusts the web page to an unknown UA.
These changes only occur after the execution of JavaScript.
Therefore, the cause of the changes is located in JS scripts on the web pages.
Hence, we can say that JS is the cause of the changes and that CSS and HTML are affected by those changes.
Looking at the content of JS scripts on the web pages, we did not find a pattern of similar instructions in the scripts.
However, the sources of the scripts produced a pattern.
76\% of the SEVERE issues were caused by third-party scripts from Google Ad Manager, while 93\% of all changes in the HTML content comparison were caused by third-party scripts from ad domains.
The remaining 7\% of domains responsible for UA changes were from bot detection and \emph{content delivery networks} (CDN) websites. 

Current ad-displaying scripts rely on known browsers, so the use of None-browsers creates invalid references.
For example, in Figure \ref{fig:severe_cases}, we showcase a prominent instance of a 'severe' problem severity case that we investigated during our study. In the first subfigure, we see the web page as accessed through a standard browser, where the rendering and layout are as intended. However, the same web page, when accessed through a None browser, as shown in the second subfigure, experiences a failure of margin collapse, a fundamental aspect of CSS layout. This failure results in a distortion in the web page's layout, as highlighted in red in the figure. Upon investigating the cause of this issue, we found a script from Google Ad Manager that was affecting the rendering of the page in None browsers. As presented in Listing \ref{margin}, the function `rn(a, b, c, d)` applies a CSS transition and a top margin to an element based on the known UA (parameter `b`). When the User Agent is not recognized, as in the case of a None browser, the function fails to apply the intended styles, causing the observed layout distortion.

\begin{figure*}[htbp]
    \centering
    \subfloat[Standard browser - normal page access]{{\includegraphics[width=8.5cm]{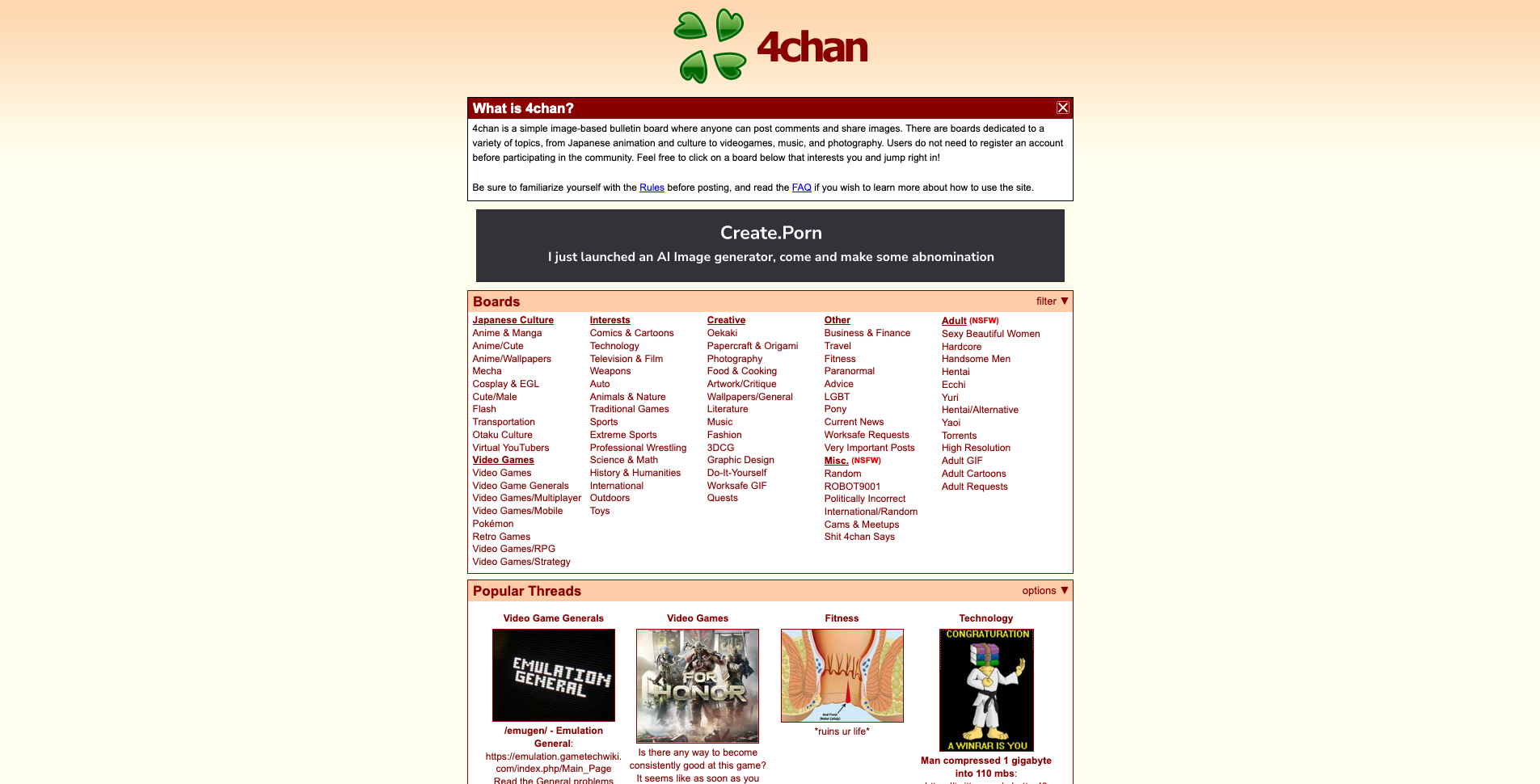}}}
    \hfill
    \subfloat[None browser - access intentionally restricted]{{\includegraphics[width=8.5cm]{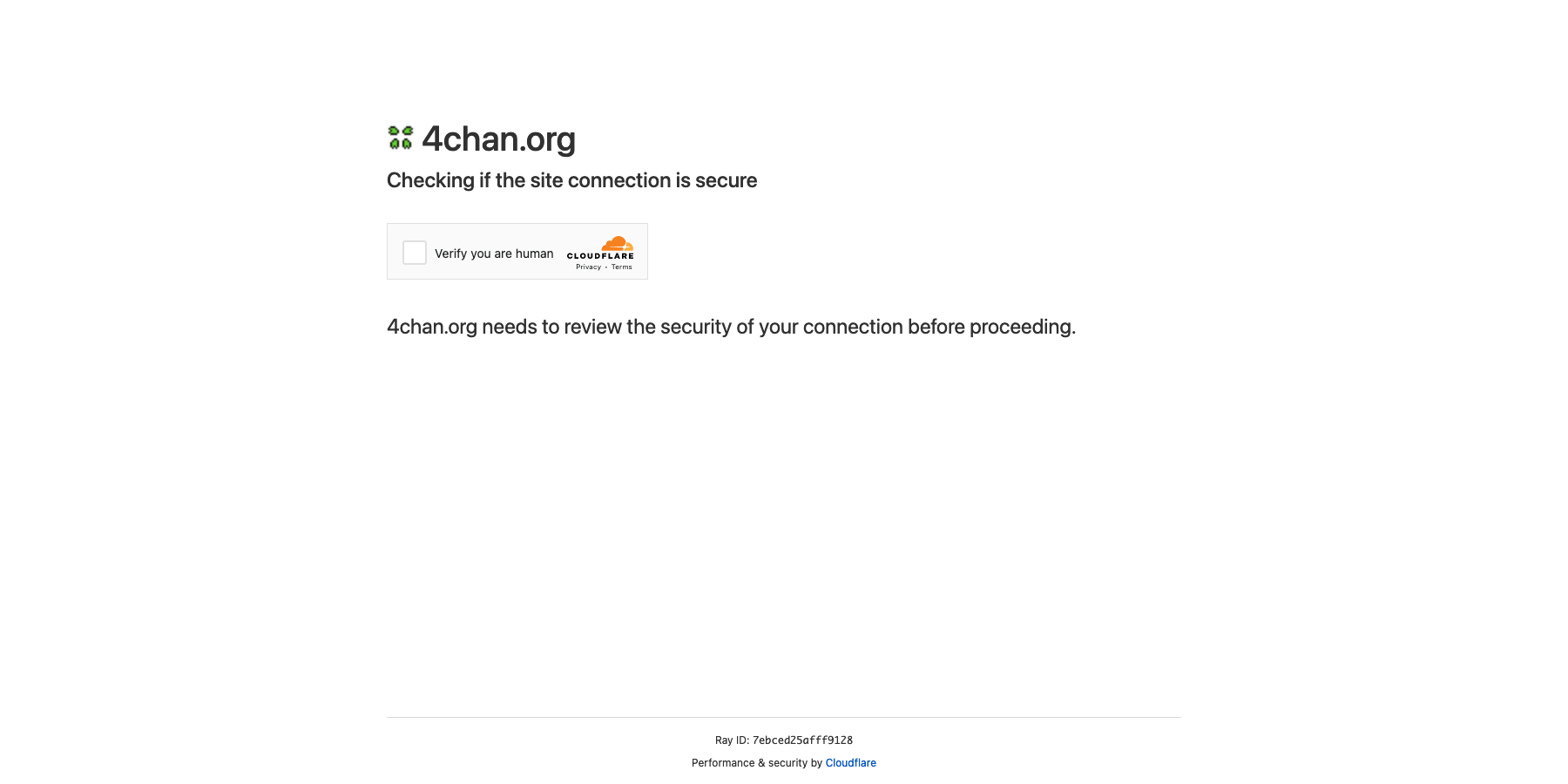}}}
    \caption{Example of "unusable" problem severity: access to the web page is intentionally restricted when using a None Browser.}
    \label{fig:unusable_cases}
\end{figure*}

\begin{lstlisting}[caption=Example of intentional restriction: a function that runs a CAPTCHA test when the UA is not recognized., label=cloudflare]
function Hl() {
  var a, b;
  return "function" === typeof(null == (a = E.navigator) ? void 0 : null == (b = a.userAgentData) ? void 0 : b.getHighEntropyValues)
}
...
Hl() ? (d(), t(r.linkAttribution)) : r.enableRecaptcha && p("require", "recaptcha", "recaptcha.js");
\end{lstlisting}

We classified such cases that cause usability issues due to code written to acknowledge the UA as an unintentional restriction. However, the cause of the remaining 7\% of the issues ranked as "UNUSABLE" was intentional. For example, In Figure \ref{fig:unusable_cases}, we illustrate a case of the "unusable" problem severity level. The first subfigure portrays a standard browser smoothly accessing a web page normally without any issues. The second subfigure illustrates a None browser attempting to access the same web page but being intentionally restricted. This is an example of the "unusable" severity level, where the user's ability to access the web page is hindered due to the intentional restriction applied when an unrecognized UA is detected. This restrictive behavior is commonly driven by JavaScript scripts embedded in the website that use the UA string to dictate access or modify the user's experience. As shown in Listing~\ref{cloudflare}, some scripts initiate a CAPTCHA test or a similar challenge when they fail to recognize the UA. In the case of a None browser, whose UA string is not recognized, this results in an intentional restriction, preventing the user from accessing the site's content.

\begin{figure}
  \centering
  \includegraphics[width=1\linewidth]{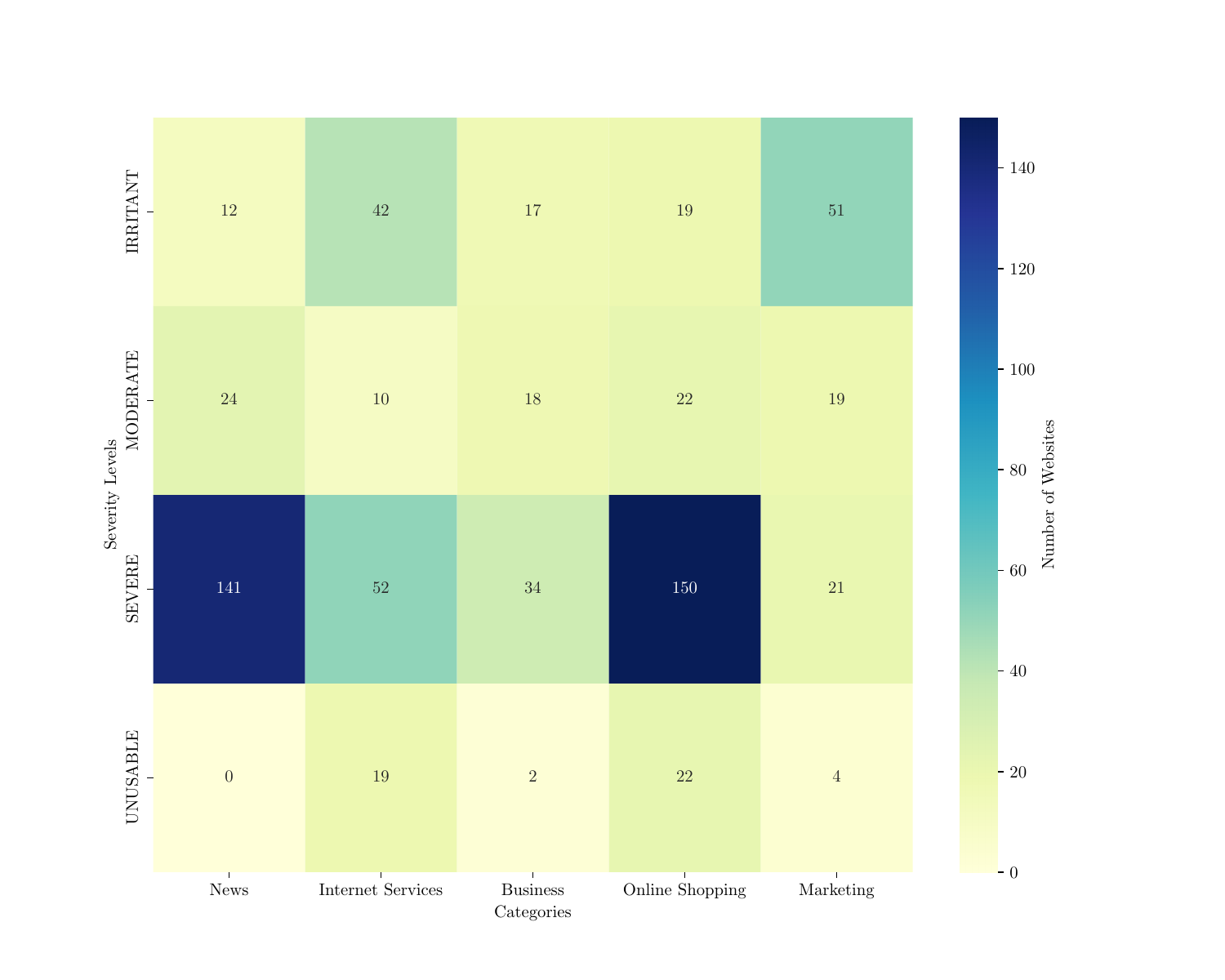}
  \caption{Problem severity distribution across website categories: this heat map depicts how changes in the UA impact different website categories, highlighting the prevalence of problem severity levels in each category.}\label{fig:severity_per_cat}
\end{figure}

In Figure \ref{fig:severity_per_cat}, we present the distribution of problem severity across the top five website categories: news, Internet services, business, online shopping, and marketing. The heat map allows us to observe the prevalence of the different severity levels, namely, irritant, moderate, severe, and unusable, across these categories.  The color intensity in each cell of the heat map is proportional to the number of websites in a category that falls under a particular problem severity level. Lighter colors indicate a lower count, while darker colors represent a higher count. The heat map provides a visual summary of our findings, revealing the extent to which different website categories are impacted by changes in the UA. For example, we can observe that the 'Severe' problem severity level is particularly prevalent in the 'news' and 'online Shopping' categories. Conversely, the 'Internet services' and 'marketing' categories have a substantial number of websites with 'irritant' or 'moderate' problem severity levels. The final observation is the pattern of the 'unusable' severity level, where the majority of occurrences are concentrated in the 'Internet services' and 'online shopping categories'.

\paragraph{\textbf{Impact of removing identifying information from the UA}} 
Firstly, the UA request-header field in the HTTP request has no impact on the web server’s response. Secondly, the {\tt navigator.userAgent} is used marginally for ads, bot detection, and CDN services, and the use of UA, in this case, causes usability problems of different severity. 
Additionally, we found that browsers could still determine that a None-browser is related to its descendant standard browser. 
The usability problems experienced due to the removal of identifying information in the UA could be fixed by adopting a feature detection approach in determining the browser in the case of bot detection or by adopting a browser-agnostic approach in writing the code in the case of ads and CDN services. 
This approach ensures that the user's privacy is protected while also promoting a secure web browsing experience.

\section{Discussion}\label{discussion}
While useful when it was introduced 3 decades ago, our study shows that the User-Agent HTTP header, which contains precise device information, has stopped being relevant on today's web. With the crawls that we performed, our results highlight that web servers do not adapt their response anymore based on the provided HTTP UA header. 
By providing different UA, all responses we collected for a single web page were identical and the only differences we observed were done by scripts that would parse the provided user agent at runtime. 
Then, the data we obtained during our crawls highlight the two following key insights:
\begin{itemize}
    \item All the standardization efforts pursued by the major web actors have had a real positive impact on the web. Browsers have become robust enough that they do not need web pages tailored for them. The browsers used on the market today implement the same set of features and provide a near-identical experience when it comes to rendering pages. 
    \item There are no major hurdles to retiring the historical HTTP User-Agent header and transitioning towards a less-granular solution like UA Client hints~\cite{ua_reduction}. As mentioned in \ref{subsec:uaweb}, the HTTP User-Agent header contributes a lot to the field of browser fingerprinting as it is one of the top attributes revealing the most information. Without it, the privacy of web users would be severely improved as there would be a lot fewer leaks of precise and unique information on the web.
\end{itemize}

\section{Impact of None-browsers on web privacy}

\begin{table}[htbp]
\centering
\begin{tabular}{l|r}
\hline
\textbf{Domain category} & \textbf{Number of Domains} \\ \hline
Total domains analyzed & $11,252$ \\
Domains accessing UA via JS API & $3,772$ \\
Domains with Vary UA-related header & $584$ \\
Domains listed on uBO & $612$ \\
UA-dependent domains & $955$ \\ \hline
\end{tabular}
\caption{Web privacy implications of UA usage: this table presents an analysis of domains based on their interaction with UA.}
\label{tab:impact_none_browsers}
\end{table}

UA strings are a critical component in various tracking techniques, including browser fingerprinting, posing a significant concern for web privacy \cite{englehardt_online_2016}. In our study, we first examined the impact of None-browsers on web page usability. To understand the potential implications on tracking techniques, we analyzed how domains in our data set accessed UA information via the JavaScript API. Specifically, 3,772 domains out of 11,252 access the UA via the JavaScript API, a common method used in browser fingerprinting \cite{nikiforakis_cookieless_2013}. Cross-referencing our dataset with a list of known trackers from uBlock Origin (uBO) \cite{hill_ublock_2023}, we found that 612 domains (5.4\%) out of the 11,252 were on the uBO list. 38.3\% of these trackers were affected by changes in the UA, suggesting that many trackers can operate without issues in the face of None-browsers or may be using other methods beyond UA strings to track users. Our findings also highlight the current state of web practices. Only 129 domains out of 11,252 contained Accept-CH response headers, suggesting that the use of Client Hints for content adaptation is not yet widespread \cite{client_hints_website}. Moreover, 584 domains return a Vary header that indicates their use of UA. Finally, our study suggests that the majority of the web remains accessible even without UA information, with only 7\% of the domains becoming unusable when browsed using a None-browser. This could encourage further adoption of None-browsers, thereby increasing user privacy.

\section{Threats to Validity}\label{threats_to_validity}
A lot of process can run on the server-side of a website and this paper focuses on the impact of UA on the client-side. 
This may threaten our conclusion on the impact of the UA on the web since the server-side is also part of the web ecosystem. 

Our conclusion on the impact of the UA on the web is also based on the fact that the None-browsers provided the string {\tt "None"} for UA and other identifying information. 
Empty, $null$ or $undefined$ UA and other identifying information may incur more breakages and other findings.

\section{Conclusion}\label{conclusion}
In this study, we investigated the role of the User Agent in today's web by crawling We crawled $270,048$ web pages from $11,252$ domains with different configurations.
Our data shows that websites no longer negotiate content based on the UA field in the HTTP request headers. 
Through JS scripts, \texttt{Navigator.userAgent} can be used for content adaptation, as few websites experience usability issues when they face an unusual user agent. 
However, the majority of those issues are unintentionally caused by third party scripts from ads, bot detection, and CDN services and can be fixed by writing browser-agnostic code. By cross-referencing our dataset with a list of known trackers from uBlock Origin, we discovered that a substantial number of known trackers did not change due to None-browsers, suggesting their robustness or the use of other tracking methods beyond UA strings.
The main takeaway of our results is that after three decades of usage, it may be finally time to retire the HTTP User Agent header and transition towards a more privacy-preserving way of sharing device information. 
The UA has been abused too many times over the years to reveal information about users and sometimes even identify them by contributing to their browser fingerprinting. 
Removing it from today's ecosystem would be a great step forward for online privacy and would contribute greatly to reducing the clutter from legacy technology.

\bibliographystyle{abbrvnat}
\bibliography{references}

\appendix

\section{Appendices}\label{appendices}

Our dataset including a full list of crawled domains and their categories, a full list of navigator properties exposed during the crawl, and other crawl and comparison data along with information on the tool we developed for \radar metrics can be found at\\\url{https://github.com/intumwa/ua-radar} 

\end{document}